\documentclass[pra,twocolumn,showpacs]{revtex4}
\usepackage{graphicx}
\usepackage{dcolumn}
\usepackage{bm}
\usepackage{amssymb}
\usepackage{amsmath}
\usepackage{epsfig}

\begin{document}

\title{Quantum Decoherence Modulated by Special Relativity}
\author{Jian-Ming Cai$^{\text{1}}$}
\author{Zheng-Wei Zhou$^{\text{1}}$}
\email{zwzhou@ustc.edu.cn}
\author{Ye-Fei Yuan$^{\text{2}}$}
\author{Guang-Can Guo$^{\text{1}}$}
\affiliation{$^{\text{1}}$Key Laboratory of Quantum Information, University of Science
and Technology of China, Chinese Academy of Sciences, Hefei, Anhui 230026,
People's Republic of China \\
$^{\text{2}}$Center for Astrophysics, University of Science and Technology
of China, Hefei, Anhui 230026, People's Republic of China}

\begin{abstract}
By investigating the evolution of a moving spin-$1/2$ Dirac electron
coupled with an a background magnetic noise, we demonstrate that the
effects of special relativity will significantly modify the
decoherence properties of the spin state. The dephasing could be
much suppressed, and for sufficiently long time the decoherence even
seems to halt. This interesting phenomenon stems from the
\textit{dressed environment} induced by special relativity.

\end{abstract}

\maketitle


\section{Introduction}
The integration of quantum mechanics and information theory gave
birth to the theory of quantum information. As another fundamental
part of modern physics theories, relativity theory also has
significant interrelationship with quantum mechanics and information
theory \cite{PTRMP,Terno}. One intriguing example is the
relativistic thermodynamics, which was renewed when quantum
properties of black holes \cite{qpbh} were discovered. The
thermodynamics of moving bodies \cite{RTMB} demonstrate that
probability distributions, which is relevant to Shannon entropy
information, depend on the inertial frame. Most recently, the
relationship of relativity theory to quantum information theory has
attracted increasing interest. Since Peres, Scudo and Terno find
that the spin Von Neumann entropy is not Lorentz invariant
\cite{PST}, the effects of Lorentz boosts on quantum states and then
quantum entanglement
\cite{Czachor,Milburn,Pachos,Adami,Terno1,Alsing,Mann,Lamata} have
been widely investigated. Most of these works about relativistic
quantum information theory (RQIT) concentrated on the quantum state
itself. RQIT maybe necessary in future practical experiments. It has
been shown that the fidelity of quantum teleportation with a
uniformly accelerated partner is reduced due to Davies-Unruh
radiation \cite{Alsing}. Other possible applications include quantum
clock synchronization \cite{Clock}, quantum-enhanced communication
\cite{EEC,Wilczewski} and global positioning \cite{global}.


Quantum decoherence is closely related to several fundamental
problems in quantum mechanics, e.g. quantum measurement and quantum
to classical transition \cite{Zurek4}. Therefore, the investigation
of decoherence in combination with special relativity is naturally
of interest and importance \cite{Breuer}, which may help gain new
insights into the fundamental issues of modern theoretic physics. It
is known that quantum systems coupled with an external environment
will suffer from inevitable decoherence, which is the most severe
obstacle to implementing quantum computation. A famous strategy to
fight against decoherence is quantum dynamical decoupling by
applying different control operations, including spin echo and
bang-bang control \cite{decoupling,spin-echo}.

Given a single spin-$1/2$ Dirac electron with the rest mass $m>0$,
we could realize the qubit by the spin up and down along the
$\hat{\mathbf{z}}$ direction. If the Dirac electron is coupled with
an noisy magnetic environment, the coherence of the spin-qubit will
be lost. In conventional quantum information theory (CQIT), people
always assume that the central quantum system is at rest. In this
paper, we present the time evolution of the spin state of a
\textit{moving} Dirac electron. It can be seen that the decoherence
properties is significantly modified by special relativity. We
demonstrate that the dressed environment induced by special
relativity will suppress the spin decoherence. One should note that,
the problem we consider here is different from the gravitational
decoherence \cite{GD}, which results from quantum metric
fluctuations and Unruh effect.

The structure of this paper is as follows. In Sec. II we demonstrate
the spin dephasing of a rest electron. In Sec. III  the influences
of special relativity on the spin decoherence properties is
investigated by presenting its dynamical behavior in the
operator-sum representation form. In Sec. IV are conclusions and
some discussions.



\section{Spin dephasing of a rest electron}
We start by considering the decoherence process for the spin degree
of freedom of an electron with a background magnetic noise. In CQIT,
the electron is always assumed to be at rest. Since the spin
magnetic moment of the electron is $\mu =\frac{e\hslash }{2mc}$,
where $e$ is the magnitude of the electronic charge, its interaction
with a magnetic field $\mathbf{B}=\nabla \times \mathbf{A}$ is
described by the following simple Hamiltonian
\begin{equation}
H_{I}=\mu \hat{\mathbf{\sigma }}\mathbf{\cdot B}
\end{equation}%
where $\hat{\mathbf{\sigma }}=(\sigma _{x},\sigma _{y},\sigma _{z})$ are
Pauli matrices. If the noisy background magnetic field is $\mathbf{B=}B\hat{%
\mathbf{z}}$ in the $\hat{\mathbf{z}}$ direction with Gaussian probability
distribution $\eta (B)=\exp (-B^{2}/2\kappa ^{2})/\sqrt{2\pi }\vartheta $,
and under quasi-static approximation \cite{Ithier}, we can write the spin
state of the electron at time $t$ as a completely positive map with an
operator-sum representation
\begin{equation}
\mathcal{E}(\rho )=\lambda _{0}(t)\rho +\lambda _{1}(t)\sigma _{z}\rho
\sigma _{z}
\end{equation}%
where $\rho $ is the initial spin state, and the parameters $\lambda
_{0}(t)+\lambda _{1}(t)=$ $1$, $\lambda _{0}(t)-\lambda _{1}(t)=$ $%
e^{-\gamma t^{2}}=\int_{-\infty }^{\infty }e^{-i2\mu Bt}\eta (B)dB$ with $%
\gamma =2\vartheta ^{2}\mu ^{2}$. Here we have set $\hbar =1$ for
simplicity. This kind of decoherence model, named dephasing, is very
important and has been widely investigated in CQIT \cite{Zurek3}. In the
dephasing process, the diagonal elements of the spin density matrix remain
unchanged. The decoherence is reflected by the off-diagonal elements, which
will decay exponentially as $\rho _{\uparrow \downarrow }(t)=\rho _{\uparrow
\downarrow }e^{-\gamma t^{2}}$ until $\rho _{\uparrow \downarrow
}(t)\rightarrow 0$ in the long time limit $\gamma t^{2}\gg 1$, i.e. the spin
state becomes classically mixed.

\begin{figure}[tbh]
\epsfig{file=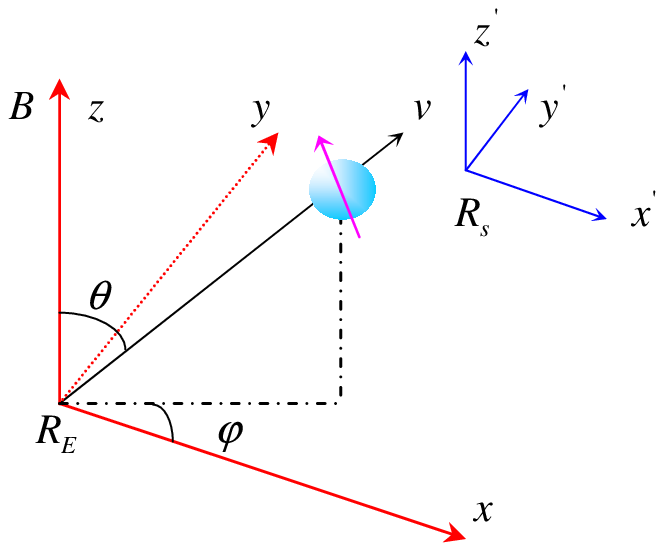,width=7cm}
\caption{(Color online) A spin-$1/2$ Dirac electron, at rest in the moving
inertial frame $R_{S}$ with the velocity $\mathbf{v}$, is coupled with the
background magnetic noise in the $\hat{\mathbf{z}}$ direction of the rest
frame $R_{E}$.}
\end{figure}

\section{Spin decoherence of a moving electron}
To deal with the relativistic Dirac electron moving at a constant
velocity $\mathbf{v}=(v\sin \theta \cos \varphi ,v\sin \theta \sin
\varphi ,v\cos \theta )$ relative to the rest frame $R_{E}$, we
should adopt the Dirac equation for the electron in external
homogeneous static fields. After choosing a suitable reference
frame $R$, the Dirac Hamiltonian in Foldy-Wouthuysen representation \cite%
{Greiner} is
\begin{eqnarray}
H_{D} &=&H_{p}+H_{SB}  \notag \\
H_{p} &=&mc^{2}+\frac{1}{2m}(\hat{\mathbf{p}}+\frac{e}{c}\mathbf{A})^{2}-%
\frac{\hat{\mathbf{p}}^{4}}{8m^{3}c^{2}} \\
H_{SB} &=&\mu \hat{\mathbf{\sigma }}\mathbf{\cdot B}-\frac{\mu }{2mc}\hat{%
\mathbf{\sigma }}\mathbf{\cdot }(\mathbf{E\times }\hat{\mathbf{p}})  \notag
\end{eqnarray}%
where $\hat{\mathbf{p}}=-i\hslash \nabla $, and $\mathbf{E=-\nabla \phi }$
represents the electric field. The above Hamiltonian is a non-relativistic
expansion to order $v_{p}^{2}/c^{2}$, where $v_{p}$ is the relative velocity
of the electron in the reference frame $R$. It is convenient for us to
investigate the dynamical properties of the Dirac electron in the the moving
inertial frame $R_{S}$ with the velocity $\mathbf{v}$ relative to the rest
frame $R_{E}$, in which the electron is at rest, i.e. $v_{p}=0$. By virtue
of the Foldy-Wouthuysen representation, we only need to consider the
positive energy states. For the spin and momentum eigen state $|\mathbf{p}%
,s\rangle $, of which $|\mathbf{p}\rangle \sim e^{i\mathbf{p\cdot r}}$, $%
H_{p}e^{i\mathbf{p\cdot r}}=\varepsilon _{p}e^{i\mathbf{p\cdot r}}$. After
time $t$, the evolution of the state is $|\mathbf{p},s\rangle \rightarrow
e^{i\phi _{p}(t)}|\mathbf{p}\rangle e^{-iH_{SB}t}|\mathbf{s}\rangle $. The
momentum phase $\phi _{p}(t)$ becomes trivial when tracing out the momentum
degree of freedom.

In the rest frame $R_{E}$, the magnetic field is $\mathbf{B=}B\hat{\mathbf{z}%
}$ in the $\hat{\mathbf{z}}$ direction, thus the fields viewed in the moving
frame $R_{S}$ can be obtained according to the Lorentz transformations \cite%
{Greiner2} as follows
\begin{eqnarray}
E_{\bot }^{\prime } &=&\cosh \xi (E_{\bot }+\frac{\mathbf{v}}{c}\times
\mathbf{B})_{\bot },\text{ \ \ }E_{\shortparallel }^{\prime
}=E_{\shortparallel }  \notag \\
B_{\bot }^{\prime } &=&\cosh \xi (B_{\bot }-\frac{\mathbf{v}}{c}\times
\mathbf{E})_{\bot },\text{ \ \ }B_{\shortparallel }^{\prime
}=B_{\shortparallel }
\end{eqnarray}%
where $\shortparallel $ and $\bot $ mean parallel and perpendicular to $%
\mathbf{v}$, the rapidity $\xi $ is defined as $\cosh \xi
=1/(1-v^{2}/c^{2})^{1/2}$. After some straightforward calculations, and note
that $\mathbf{p=0}$, we get the effective Hamiltonian for the spin degree of
freedom of the Dirac electron $H_{SB}=\mu \hat{\mathbf{\sigma }}\mathbf{%
\cdot B}^{\prime }$, where $B_{x}^{\prime }=B(1-\cosh \xi )\cos \theta \sin
\theta \cos \varphi $, $B_{y}^{^{\prime }}=B(1-\cosh \xi )\cos \theta \sin
\theta \sin \varphi $, and $B_{z}^{^{\prime }}=B(\cos ^{2}\theta +\cosh \xi
\sin ^{2}\theta )$. The above effective interaction Hamiltonian means that
from the viewpoint of the moving Dirac electron, the environment is
different compared to the situation when the electron is at rest due to the
relativity of the magnetic field. In the following, we will investigate in
detail how this kind of effects will modify the dynamical properties of the
spin decoherence.

The interaction with the external magnetic field $\mathbf{B}^{\prime }$ will
introduce a rotation on the spin by $\delta $ about the $\hat{\mathbf{n}}%
=(n_{x},n_{y},n_{z})$ axis as
\begin{equation}
U(B,t)=\exp (-i\frac{\delta }{2}\hat{\mathbf{\sigma }}\mathbf{\cdot }\hat{%
\mathbf{n}})  \label{RT}
\end{equation}%
where $n_{i}=B_{i}^{\prime }/B^{^{\prime }}$, $i=x,y,z$, with $B^{^{\prime
}}=(B_{x}^{\prime 2}+B_{y}^{\prime 2}+B_{z}^{\prime 2})^{1/2}=\kappa B$, $%
\kappa =(\cos ^{2}\theta +\cosh ^{2}\xi \sin ^{2}\theta )^{1/2}$, and the
rotation angle is $\delta =2\mu tB^{^{\prime }}=2\kappa \mu tB$. In the
similar way, the noisy background magnetic field $B$ is with Gaussian
probability distribution, and under quasi-static approximation \cite{Ithier}%
, we can write the spin state of the moving Dirac electron at time $t$ as%
\begin{equation}
\rho (t)=\int_{-\infty }^{\infty }\rho (B,t)\eta (B)dB
\end{equation}%
where $\rho (B,t)=U(B,t)\rho U^{\dag }(B,t)$ and $\rho $ denotes the initial
spin state.

We first examine the diagonal elements by calculating $\rho _{\uparrow
\uparrow }(t)$. According to Eq.(\ref{RT}), it is easy for us to write $\rho
_{\uparrow \uparrow }(B,t)=\rho _{\uparrow \uparrow }-2\Delta \rho
_{\uparrow \uparrow }\sin ^{2}\frac{\delta }{2}+\frac{i}{2}(\rho _{\uparrow
\downarrow }e^{i\varphi }-\rho _{\downarrow \uparrow }e^{-i\varphi
})(n_{x}^{2}+n_{y}^{2})^{1/2}\sin \delta $, where $\Delta \rho _{\uparrow
\uparrow }=\frac{1}{2}[\eta (\rho _{\uparrow \uparrow }-\rho _{\downarrow
\downarrow })-\chi (\rho _{\uparrow \downarrow }e^{i\varphi }+\rho
_{\downarrow \uparrow }e^{-i\varphi })]$ with two modulation factors defined
as $\eta =(1-n_{z}^{2})$ and $\chi =n_{z}(n_{x}^{2}+n_{y}^{2})^{1/2}$. After
integrating over the magnetic field $B$ with Gaussian probability
distribution, we obtain
\begin{equation}
\rho _{\uparrow \uparrow }(t)=\rho _{\uparrow \uparrow }-\Delta \rho
_{\uparrow \uparrow }(1-e^{-\gamma ^{\prime }t^{2}})  \label{DE}
\end{equation}%
where $\gamma ^{\prime }=\kappa ^{2}\gamma =2\kappa ^{2}\vartheta ^{2}\mu
^{2}$. The first term is just the same as the dephasing process in CQIT,
i.e. $v=0$. However, the second item in Eq.(\ref{DE}) indicates that the
diagonal elements will change as time, which is a different decoherence
source introduced by the effects of special relativity. In the long time
limit $\gamma ^{\prime }t^{2}\gg 1$, the spin up population will decrease $%
\Delta \rho _{\uparrow \uparrow }$.

Now we turn to the off-diagonal elements, in the similar way we can obtain $%
\rho _{\uparrow \downarrow }(B,t)=\rho _{\uparrow \downarrow }e^{-i\delta
}+2\Delta \rho _{\uparrow \downarrow }\sin ^{2}\frac{\delta }{2}+\frac{i}{2}%
[(\rho _{\uparrow \uparrow }-\rho _{\downarrow \downarrow
})(n_{x}-in_{y})+2\rho _{\uparrow \downarrow }(1-n_{z})]\sin \delta $, where
$\Delta \rho _{\uparrow \downarrow }=\frac{1}{2}[\chi (\rho _{\uparrow
\uparrow }-\rho _{\downarrow \downarrow })e^{-i\varphi }+\eta (\rho
_{\uparrow \downarrow }+\rho _{\downarrow \uparrow }e^{-i2\varphi })]$.
After integrating over the Gaussian magnetic field $B$, we obtain
\begin{equation}
\rho _{\uparrow \downarrow }(t)=\rho _{\uparrow \downarrow }e^{-\gamma
^{\prime }t^{2}}+\Delta \rho _{\uparrow \downarrow }(1-e^{-\gamma ^{\prime
}t^{2}})  \label{NDE}
\end{equation}%
The first term of $\rho _{\uparrow \downarrow }(t)$ is in a similar form of
exponential decay, except that the decay rate changes to $\gamma ^{\prime
}=\kappa ^{2}\gamma \geq \gamma $. The effects of special relativity is also
reflected by the second term, which implies that when $\gamma ^{\prime
}t^{2}\gg 1$, $\rho _{\uparrow \downarrow }(t)\rightarrow \Delta \rho
_{\uparrow \downarrow }$ that is a saturation value other than zero, i.e.
the off-diagonals will not vanish.

From the evolution of diagonal and off-diagonal elements in Eqs.(\ref{DE},%
\ref{NDE}), we can establish a physical picture of the above analyses by
expressing the spin density matrix at time $t$ in the operator-sum
representation%
\begin{equation}
\mathcal{E}_{m}(\rho )=p_{0}\rho +p_{1}\sigma _{z}\rho \sigma
_{z}-\varepsilon \sigma _{z}\rho \sigma _{z}+\sum\limits_{i=1}^{2}F_{i}\rho
F_{i}^{\dag }  \label{OSR}
\end{equation}
where $p_{0}=(1+e^{-\gamma ^{\prime }t^{2}})/2$, $p_{1}=(1-e^{-\gamma
^{\prime }t^{2}})/2$ and $\varepsilon =p_{1}(\eta +\chi )$. The operators $%
\left\{ F_{i}\right\} $ are $F_{1}=[p_{1}(\eta -\chi )]^{1/2}(\cos
\varphi \sigma _{x}+\sin \varphi \sigma _{y})$ and $F_{2}=(p_{1}\chi
)^{1/2}(\sigma _{z}+\cos \varphi \sigma _{x}+\sin \varphi \sigma
_{y})$. If the velocity of the electron $v=0$, the modulation factor
$\eta =\chi =0$, and the above results reduce to the pure dephasing
of a rest spin. When we consider a moving electron, the first two
terms in Eq.(\ref{OSR}) is similar to pure dephasing. However, the
decay rate of the off-diagonal elements is amplified by the factor
$\kappa ^{2}\geq 1$. The suppression of dephasing stems from the
third term $-\varepsilon \sigma _{z}\rho \sigma _{z}$. The other two
operators $F_{1}$ and $F_{2}$ in Eq.(\ref{OSR}) represent different
decoherence mechanisms which makes the spin suffer from other
decoherence than pure dephasing. We could also formulate the
evolution of the spin state of the moving electron as in the dressed
environment
\begin{widetext}
\begin{equation}
\rho (t)=V^{\dagger }\left\{ \int_{-\infty }^{\infty }[\exp (-i\frac{\delta
_{0}}{2}\sigma _{z})]^{\kappa }(V\rho V^{\dagger })[\exp (i\frac{\delta _{0}%
}{2}\sigma _{z})]^{\kappa }\eta (B)dB\right\} V
\end{equation}
\end{widetext}
where $V=\exp [-i\frac{\phi }{2}\hat{\mathbf{\sigma }}\mathbf{\cdot }(\hat{%
\mathbf{n}}\times \hat{\mathbf{z}})]$ is the dressing
transformation, $\phi $ is the angle between the axes
$\hat{\mathbf{n}}$ and $\hat{\mathbf{z}}$. From the viewpoint of
qubit, after the control operation $V$, the initial state $\rho$
will undergo pure dephasing in the dressed Hilbert space, the final
reverse operation $V^{\dagger}$ will recover some coherence
information in the original Hilbert space.

\begin{figure}[tbh]
\epsfig{file=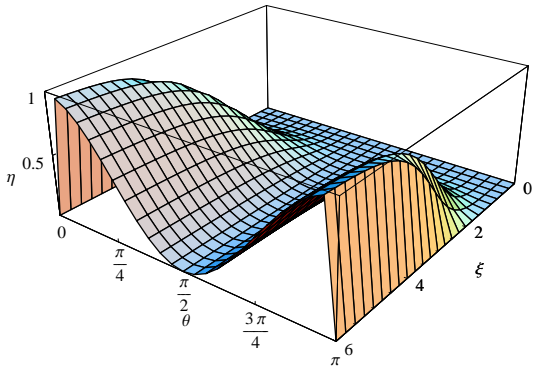,width=7cm}
\caption{{}(Color online) Decoherence modulation factor as a function of the
moving rapidity and angle: $\protect\eta $\textit{\ vs. }$\protect\xi $ and $%
\protect\theta $. }
\end{figure}

The modulation factors $\eta $ and $\chi $ are the key ingredients
that characterize the influence of special relativity. Fig 2 shows
the modulation factor $\eta $ as a function of the rapidity $\xi $
and $\theta $. For a given rapidity $\xi $, the modulation factor
$\eta =(\cosh \xi -1)^{2}(1-\cos ^{2}2\theta )/2[(\cosh ^{2}\xi
+1)-(\cosh ^{2}\xi -1)\cos
2\theta ]$. We are interested in the maximum value of the modulation factor $%
\eta $. Thus we consider the first order partial differential equation $%
\partial \eta /\partial (\cos 2\theta )=0$, which leads to $\cos 2\theta
=(\cosh \xi -1)/(\cosh \xi +1)$, and the corresponding maximum value of $%
\eta $ is%
\begin{equation}
\eta _{\max }=(\frac{\cosh \xi -1}{\cosh \xi +1})^{2}
\end{equation}%
We plot the maximum value $\eta _{\max }$ for various rapidity $\xi $ in the
following Fig 3(a). It can be seen that $\eta _{\max }$ always increases
monotonically as the rapidity $\xi $ grows. In the limit $\xi \rightarrow
\infty $, i.e. the velocity is close to the light velocity $v\rightarrow c$,
the maximum modulation factor $\eta _{\max }$ reaches the constant value $1$%
.
\begin{figure}[tbh]
\epsfig{file=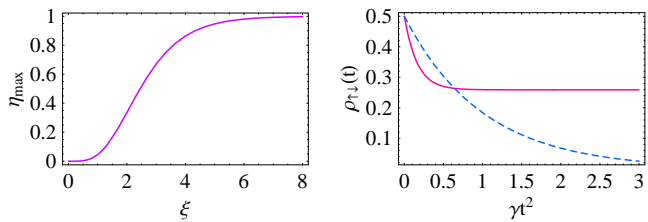,width=8.5cm}
\caption{(Color online) \textit{a.} Maximum value of decoherence modulation
factor as a function of rapidity: $\protect\eta _{\max }$ \emph{vs.} $%
\protect\xi $; \textit{b. }Off-diagonal element as a function of time: $%
\protect\rho _{\uparrow \downarrow }(t)$ \textit{vs. }$\protect\gamma t^{2}$%
, and the rapidity $\protect\xi =2.5$ (Solid), $\protect\xi =0$ (Dashed).
The initial spin state is $|\protect\psi \rangle =\frac{1}{\protect\sqrt{2}}%
(|\uparrow \rangle +|\downarrow \rangle )$.}
\end{figure}
For the angle $\theta $ of the velocity $v$ that maximize the value of $\eta
$, the corresponding value of the other modulation factor $\chi
=n_{z}(n_{x}^{2}+n_{y}^{2})^{1/2}$ is $\chi _{m}=2\cosh ^{1/2}\xi (\cosh \xi
-1)/(\cosh \xi +1)^{2}$. When we consider the relativistic Dirac electron,
i.e. the velocity $v$ is comparable to the light velocity $c$, $\chi _{m}\ll
\eta _{\max }$, which can be neglected in some sense.

\emph{Example }To explicitly demonstrate the influence of special relativity
on the spin decoherence of a moving Dirac electron, we consider a spin qubit
in a fully coherent initial state $|\psi \rangle =\frac{1}{\sqrt{2}}%
(|\uparrow \rangle +|\downarrow \rangle )$. We could have chosen a more
general initial spin state with different relative amplitudes. However, the
above state gives rise to all interesting physical features in the situation
considered here.
We set the velocity angle $\varphi =0$ to maximize the modulus of the
saturation value of off-diagonal elements, thus
\begin{eqnarray}
\rho _{\uparrow \uparrow }(t) &=&\frac{1}{2}[1+\chi (1-e^{-\gamma ^{\prime
}t^{2}})] \\
\rho _{\uparrow \downarrow }(t) &=&\frac{1}{2}[(1-\eta )e^{-\gamma ^{\prime
}t^{2}}+\eta ]
\end{eqnarray}%
We plot the decay of the off-diagonal elements of the spin density matrix in
Fig 3(b), and compare two cases when the rapidity $\xi =2.5$ and $\xi =0$.
The velocity angle $\theta $ is set to maximize the modulation factor $\eta $%
. 
In the short time region, the off-diagonal elements decay much faster, which
is due to the amplification of $\gamma \rightarrow \gamma ^{\prime }\ $by
the factor $\kappa ^{2}=6.13229>1$. However, as the time is longer, the
decay of the off-diagonal elements will be much suppressed. In particular,
if the time is sufficiently long, for $\xi =0$, the spin degree of freedom
becomes completely classical; while for $\xi =2.5$, the off-diagonal
elements will reach a nonzero saturation value $\eta _{\max }/2=0.2589$, and
the spin state becomes steady, which suggests that the decoherence process
seems halting.


The above discussions are extensible to the situation of several
spin-$1/2$ Dirac electrons coupled with one common Gaussian
background magnetic noise, i.e.
$H_{SB}=\mu\sum\limits_{i}\hat{\mathbf{\sigma}}^{i}\mathbf{\cdot B}$
\cite{Braun,cai2}. For example, we consider two electrons in the spin entangled state $%
|\psi\rangle =\frac{1}{\sqrt{2}}(|\uparrow\uparrow \rangle +|\downarrow
\downarrow \rangle )$ moving at the same velocity,
if the electrons are at rest, the two-qubit entanglement quantified
by the concurrence will decay exponentially as $\mathcal{C}(t)=\exp
(-4\gamma t^{2}) $ \cite{Lidar,cai}. In order to highlight the
effects of special relativity, we assume that $v\cong c$, thus the
modulation factors are $\eta \cong 1$ and $\chi \cong 0$. After
simple calculations, it is easy to obtain the evolution of the
two-qubit entanglement as $\mathcal{C}^{\prime }(t)=\exp (-4\gamma
^{\prime }t^{2})$. Therefore, unlike the situation of a
single electron, the effects of special relativity will makes $\mathcal{C}%
^{\prime }(t)\rightarrow 0$ much more quickly.

\section{Conclusions}
The existing research in the field of RQIT focused on the effects of
special relativity on the \textit{static} properties of quantum
states, e.g. Von Neumann entropy and quantum entanglement. In this
work, we investigate the \textit{dynamic} properties, i.e.
decoherence process of a moving Dirac electron. The decoherence
mechanisms will be significantly modified due to the \textit{dressed
environment}, which leads to the intriguing phenomenon that the
coherent information will be preserved even as the electron is in
the noisy environment for sufficiently long time.
The extension of this work to general decoherence models will
enlarge the research scope of relativistic quantum information
theory, and may establish fundamental connections between special
relativity and quantum mechanics.


\section{Acknowledgments}
We thank Prof. Daniel A. Lidar and L. Lamata for valuable
discussions. This work was funded by National Fundamental Research
Program, NCET-04-0587, the Innovation funds from Chinese Academy of
Sciences, and National Natural Science Foundation of China (Grant
No. 60121503, 10574126).

\end{document}